\documentclass[aps,showpacs,manuscript,12pt]{revtex4}
\usepackage{amssymb}
\usepackage{amsmath}
\usepackage{graphicx}

\begin{document}

\title{\textbf{On the existence of canonical gyrokinetic variables for
chaotic magnetic fields$^{\S}$}}
\author{Piero Nicolini$^{a,b}$ and Massimo Tessarotto$^{a,b}$}
\affiliation{\ $^{a}$Department of Mathematics and Informatics,
University of Trieste, Italy\\ $^{b}$Consortium of
Magneto-fluid-dynamics, University of Trieste, Italy}

\begin{abstract}
The gyrokinetic description of particle dynamics faces a basic
difficulty when a special type of canonical variables is sought,
i.e., the so-called \textit{gyrokinetic canonical variables}.
These are defined in such a way that two of them are respectively
identified with the gyrophase-angle, describing the fast particle
gyration motion around magnetic field lines, and its canonically
conjugate momentum. In this paper we intend to discuss the
conditions of existence for these variables.
\end{abstract}

\pacs{52.30.Gz,52.30.-q}
\date{\today }
\maketitle



\section{Introduction}

The gyrokinetic description of particle dynamics concerns the
representation - in terms of generally non-canonical variables -
of the state of a single classical charged point-particle immersed
in an electromagnetic (EM) field. In classical electrodynamics -
for arbitrary EM fields and provided the EM self-force is
neglected (according to the customary interpretation; see,
however, the related discussion in Ref. \cite{Dorigo2008a}) - it
is well known that this defines an Hamiltonian dynamical system.
As such, locally in phase-space, it can always be represented in
canonical form. Nevertheless, although well-known in the
literature, the gyrokinetic problem presents a basic difficulty
related specifically to the construction of a special set of
canonical variables. As described below, this arises in connection
with the construction of the so-called \emph{canonical gyrokinetic
variables}. \ In this paper we intend to show that, unless special
symmetry conditions hold (in particular for the magnetic field),
such variables do not exist.

For a proper formulation of the problem it is important to stress that the
gyrokinetic description of particle dynamics concerns in principle two
possible viewpoints :

\emph{A) an exact representation (of the particle state) realized by
suitably prescribing an appropriate phase-space diffeomorphism [see below
Eq.(\ref{1})];}

\emph{B) an approximate representation, obtained by means of a suitable
asymptotic expansion. }

In the exact gyrokinetic treatment (Approach A) it is assumed that there
exists a phase-space mapping (i.e., a diffeomorphism, which is assumed to
exist at least in a suitable subset of phase-space) of the form%
\begin{equation}
(\mathbf{r,v})\rightarrow (\mathbf{y}^{\prime },\phi ^{\prime })  \label{1}
\end{equation}%
between the Newtonian particle state $(\mathbf{r,v})$ (with $\mathbf{r}$ and
$\mathbf{v}$ denoting respectively the particle position and velocity) and
the (real) gyrokinetic state $(\mathbf{y}^{\prime },\phi ^{\prime }).$ The
latter is defined in such a way that equations of motion for the transformed
variables take the form

\begin{equation}
\left\{
\begin{array}{c}
\frac{d\mathbf{y}^{\prime }}{dy}=\mathbf{Y(y}^{\prime },t), \\
\frac{d\phi ^{\prime }}{dt}=F(\mathbf{y}^{\prime },t).%
\end{array}%
\right.
\end{equation}%
By construction $\mathbf{Y(y}^{\prime },t)$ and $F(\mathbf{y}^{\prime },t),$
to be identified with suitably smooth real functions of the gyrokinetic
state, are both assumed independent of the variable $\phi ^{\prime }$.
Hence, its canonically conjugate momentum $p_{\phi ^{\prime }}$ is
necessarily a first integral of motion. \ Here $\phi ^{\prime }$ and $%
\mathbf{y}^{\prime }$ are the so-called gyrokinetic (or guiding-center)
variables, representing respectively an angle (the so-called gyrophase)
which describes the particle gyration motion around the magnetic flux lines
and an arbitrary 5-dimensional real vector, representing a \textit{reduced
non-canonical gyrokinetic state}. As an example, in particular, the vector $%
\mathbf{y}^{\prime }$ may be identified with the non-canonical variables
\begin{equation}
\mathbf{y}^{\prime }\equiv \left( \mathbf{r}^{\prime },\xi ^{\prime
},p_{\phi ^{\prime }}\right) ,  \label{non-canonical
reduced state}
\end{equation}%
where $\mathbf{r}^{\prime }$ and $\xi ^{\prime }$ \ denote respectively the
guiding-center position vector and an additional (independent)
velocity-space gyrokinetic variable.

However, unless the electromagnetic field is specially prescribed (i.e., for
example, it is a constant), the exact gyrokinetic transformation (\ref{1})\
cannot generally be achieved. Nevertheless, it is still possible - under
suitable assumptions - to determine it in an approximate sense by means of
an appropriate asymptotic approximation for the (canonical or non-canonical)
particle state (Approach B). This is obtained by introducing an asymptotic
expansion (i.e., a truncated perturbative expansion in terms of an
appropriate infinitesimal dimensionless parameter $\varepsilon $) of the form%
\begin{equation}
(\mathbf{r,v})\rightarrow (\mathbf{y}^{\prime },\phi ^{\prime })\cong
\sum\limits_{k=0}^{N}\varepsilon ^{k}(\mathbf{y}_{k}^{\prime },\phi
_{k}^{\prime })+o(\varepsilon ^{N+1})  \label{1bis}
\end{equation}%
where the integer $N>1$ (to be suitably prescribed) denotes the "order" of
the asymptotic approximation. In particular, by assuming that the magnetic
field in which the particle is immersed is suitably \textquotedblleft
intense\textquotedblright , \ the infinitesimal dimensionless parameter can
be defined as%
\begin{equation}
\varepsilon =r_{L}/L<<1.  \label{2}
\end{equation}%
Here the notation is standard. \ Thus, $L$ and $r_{L}$ are respectively a
characteristic scale length of the EM fields (to be suitably defined, see
below) and the velocity-dependent particle Larmor radius $r_{L}=\frac{%
w^{\prime }}{\Omega ^{\prime }}.$ In particular, all primed quantities are
evaluated at the guiding-center position $\mathbf{r}^{\prime },$ which
requires that the diffeomorphism
\begin{equation}
\mathbf{r\rightarrow r}^{\prime }\cong \sum\limits_{k=0}^{N}\varepsilon ^{k}%
\mathbf{r}_{k}^{\prime }+o(\varepsilon ^{N+1})  \label{1ter}
\end{equation}%
is assumed to exist. For example $\Omega _{\text{ }}^{\prime }\equiv \Omega
_{\text{ }}^{\prime }(r^{\prime },t),$ where $\Omega _{\text{ }}^{\prime }=%
\frac{qB^{\prime }}{mc}$ is the Larmor frequency and $q,m,B$ are
respectively the charge and mass of a point particle and the magnitude of
the magnetic field. Moreover, $\mathbf{w}^{\prime }$ is the orthogonal
component of the particle velocity to be evaluated - in a suitable reference
frame - at the same position $\mathbf{r}^{\prime }.$

The first author who systematically investigated the gyrokinetic problem,
based on the explicit construction of an asymptotic expansion of the form (%
\ref{1bis}) and (\ref{1ter}), was Alfven \cite{Alfen 1950} who pointed out
the existence of an adiabatic invariant, the magnetic moment $\mu ^{\prime
}\equiv \frac{qc}{m}$ $p_{\phi }^{\prime },$ in the sense:
\begin{equation}
\frac{d}{dt}\ln \mu ^{\prime }\sim O(\varepsilon ).
\end{equation}%
After subsequent work which dealt with direct construction methods of
gyrokinetic variables \cite{Gardner 1959,Northrop et al. 1960,Morozov 1966},
a significant step forward was made by Kruskal \cite{Kruskal 1962} who,
first, established the consistency of the Alfven approach by proving, under
suitable assumptions on the EM fields, that the magnetic moment can be
constructed correct at any order $N$ in $\varepsilon $ in such a way that,
denoting $M^{\prime }$ such a dynamical variable, it is an adiabatic
invariant of order $N,$ namely in the sense%
\begin{eqnarray}
\frac{d}{dt}\ln M^{\prime } &\sim &O(\varepsilon ^{N}) \\
M^{\prime } &=&\mu ^{\prime }+\varepsilon \mu _{1}^{\prime }+...+\varepsilon
^{N}\mu _{N}^{\prime }.
\end{eqnarray}%
A modern picture of the Hamiltonian formulation, which makes easier the
formulation of higher order perturbative theories, was given only later in
terms of Lie-transform methods \cite{Cary et al. 1977,Johnston et al.
1978,Littlejohn1979}. However, it was only with the adoption of
non-canonical Lie-transform methods \cite{Littlejohn1979} that the approach
was given a general formulation. As a motivation to his non-canonical
approach, Littlejohn \cite{Littlejohn1979,Littlejohn1981} pointed out what
in his views was a critical point of purely canonical formulations such as
previously developed Lie transform approaches \cite{Cary et al.
1977,Johnston et al. 1978}, namely the ambiguity in the separation of the
unperturbed and perturbed contributions in the Hamiltonian due the presence
of the vector potential $\mathbf{A}$ in the canonical momenta. He showed
that this difficulty can be circumvented by making use of suitable
non-canonical variables which include the canonical pair $(\phi ^{\prime
},p_{\phi }^{\prime }).$

The possibility of constructing canonical gyrokinetic variables has relied,
since, on two possible methods

\begin{itemize}
\item the Darboux reduction algorithm, based first on the construction of a
set of non-canonical gyrokinetic variables \cite{Littlejohn1983};

\item the direct construction of canonical gyrokinetic variables, either in
terms of mixed-variables generating functions \cite{Gardner 1959}, canonical
Lie-transform methods \cite{Cary et al. 1977,Johnston et al.
1978,Tessarotto2004} or based on the adoption of the hybrid Hamilton
variational principle \cite{Littlejohn1983,Pozzo,Beklemishev1999}.
\end{itemize}

The first approach, and probably the most popular in the literature \cite%
{White et al. 1984,Hahm Lee Brizard 1988,Hahm 1988,White 1990}, is the based
on the use of Darboux theorem which allows, in principle, the construction
of canonical variables for an arbitrary differential 1-form. The canonical
1-form expressed in terms of the canonical variables is then obtained by
applying recursively the so-called \textquotedblleft Darboux reduction
algorithm\textquotedblright as pointed out by Littlejohn, which is obtained
by a suitable combination of dynamical gauge and coordinate transformations.
Nevertheless this approach \ leads to potential complications and
ambiguities due to the fact that the so-called canonical gyrokinetic
coordinates (see below) are field-related. Therefore it would be highly
desirable to be able to construct gyrokinetic canonical variables which
result independent of the magnetic field geometry. In a previous work {%
Tessarotto and Nicolini, 2006 \cite{Tessarotto2006}) a possible solution to
this problem has been pointed out by adopting superabundant canonical
variables. Purpose of this work is - instead - to address the problem of the
construction of \textit{essential }}gyrokinetic canonical variables.

\section{Lagrangian approach and canonical gyrokinetic reduction}

Starting point for the application of the Darboux reduction method is the
standard Lagrangian formulation for gyrokinetic particle dynamics, expressed
in non-canonical gyrokinetic variables. For definiteness, let us briefly
recall its formulation. Let us assume for this purpose that the EM
potentials $\Phi ,\ \mathbf{A}$ \ are analytic functions of $\varepsilon $
and hence can be represented in power series of $\varepsilon $%
\begin{eqnarray}
\Phi  &=&\sum_{i=-1}^{N}\varepsilon ^{i}\Phi _{i}(\mathbf{r},t), \\
\mathbf{A} &=&\sum_{i=-1}^{N}\varepsilon ^{i}\mathbf{A}_{i}(\mathbf{r},t).
\end{eqnarray}%
Here $\varepsilon $ is the infinitesimal dimensionless parameter defined
above (\ref{1}). In particular, the characteristic scale length $L$ entering
its definition is identified with the minimum of the gradient-scale lengths
for the perturbations of the EM potentials $\left( \Phi _{i},\mathbf{A}%
_{i}\right) $, namely as $L\leq \min \left( \frac{1}{\left\vert \mathbf{A}%
_{i}\right\vert }\left\vert \frac{\partial \left\vert \mathbf{A}%
_{i}\right\vert }{\partial \mathbf{r}}\right\vert \right) ^{-1},$ $\min
\left( \frac{1}{\left\vert \Phi _{i}\right\vert }\left\vert \frac{\partial
\Phi _{i}}{\partial \mathbf{r}}\right\vert \right) ^{-1}$ for $i=-1,N$. In
addition, denoting where $\mathbf{b}(\mathbf{r},t)\mathbf{=B}(\mathbf{r}%
,t)/B(\mathbf{r},t),$ the magnitudes of the particle velocity $\left\vert
\mathbf{v}\right\vert $ and of the electric drift velocity $\mathbf{v}_{E}=c%
\mathbf{E}\times \mathbf{b}/B$ are assumed of the same order, in the sense
\begin{equation}
\left\vert \mathbf{v}\right\vert /\left\vert \mathbf{v}_{E}^{\prime
}\right\vert \sim o(1)
\end{equation}%
and consequently the parallel electric field is similarly ordered as
\begin{equation}
\mathbf{b}\cdot \mathbf{E}\sim o(1)  \label{small parallel E
field}
\end{equation}%
(\textit{condition of small parallel electric field}). In validity of these
hypotheses the construction of the standard gyrokinetic variables is well
known and has been achieved by several authors (see for example \cite%
{Littlejohn1979}). For definiteness, let us identify the reduced gyrokinetic
state $\mathbf{y}^{\prime }$ with $\mathbf{y}^{\prime }=\left( \mathbf{r}%
^{\prime },u^{\prime },p_{\phi }^{\prime }\right) $. Here $u^{\prime }$
denotes the parallel velocity%
\begin{equation}
u^{\prime }=\mathbf{b}^{\prime }\cdot \left( \mathbf{v}^{\prime }\mathbf{-v}%
_{E}^{\prime }\right) ,
\end{equation}%
where $\mathbf{v}^{\prime }$ is the guiding-center velocity and $\mathbf{v}%
_{E}^{\prime }=-\frac{c}{B^{\prime }}\mathbf{b}^{\prime }\times \nabla
^{\prime }\phi ^{\prime }$ is the $\mathbf{E\times B}$-drift velocity
evaluated at the guiding-center position. In this case - and in the presence
of slowly varying EM fields - the fundamental Lagrangian differential 1-form
expressed in terms of gyrokinetic variables reads
\begin{equation}
\left. d\Gamma ^{\prime }\equiv dt\mathcal{L}^{\prime }(\mathbf{y}^{\prime },%
\mathbf{v}^{\prime },\dot{{\mathbf{r}}}^{\prime },\dot{{\phi }}^{\prime
},t)=dG^{\prime }-d{\phi }^{\prime }p_{\phi }^{\prime }-dtH^{\prime }\right.
\label{Lagrangan 1-form}
\end{equation}%
where $dG^{\prime }$ and $H^{\prime }$ are respectively the exchange term
\begin{eqnarray}
dG^{\prime } &\equiv &\mathbf{a(y}^{\prime },t)\cdot d{\mathbf{r}}^{\prime },
\label{exchange term} \\
\mathbf{a(y}^{\prime },t) &\equiv &\frac{q}{\varepsilon c}\mathbf{A}^{\ast }(%
\mathbf{y}^{\prime },t),
\end{eqnarray}%
and the gyrokinetic Hamiltonian
\begin{equation}
H^{\prime }\equiv \frac{m}{2}\mathbf{v}^{\prime 2}+\mu ^{\prime }B^{\prime }+%
\frac{q}{\varepsilon }\Phi ^{\ast }(\mathbf{y}^{\prime },t).  \label{GKT HAM}
\end{equation}%
Moreover, for definiteness, let us identify the reduced state $\mathbf{y}%
^{\prime }$ with a suitable "effective" vector potential $%
\mathbf{A}^{\ast }(\mathbf{y}^{\prime },t),$ i.e., it reads $.$ Here, both
the exchange term and gyrokinetic Hamiltonian, in particular the effective
EM potentials $\left( \Phi ^{\ast },\mathbf{A}^{\ast }\right) $ are
expressed as functions only of the non-canonical reduced gyrokinetic state $%
\mathbf{y}^{\prime }$ defined above (\ref{non-canonical reduced state}). \
In the following the gyrokinetic differential 1-form $d\Gamma ^{\prime }$
will be considered either exactly prescribed (Approach A) or determined in
terms of an asymptotic approximation of order $o(\varepsilon ^{N+1})$,
namely neglecting corrections of order $o(\varepsilon ^{N+1})$ to $d\Gamma
^{\prime }$ (Approach B).

Let us now seek a diffeomorphism, to be assumed at least locally defined in
the relevant phase-space, of the form
\begin{equation}
\left( \mathbf{r}^{\prime },u^{\prime },p_{\phi ^{\prime }},\phi ^{\prime
}\right) \rightarrow \left( q^{\prime 1},q^{\prime 2},p_{1}^{^{\prime
}},p_{2}^{\prime },p_{\phi ^{\prime }},\phi ^{\prime }\right) ,
\label{canonical gyrokinetic variables}
\end{equation}%
where $y^{\prime 1},y^{\prime 2},p_{1}^{^{\prime }}$ and $p_{2}^{\prime }$
are assumed smooth real functions only of $\mathbf{y}^{\prime }$ (and hence
by definition as gyrokinetic variables). Provided the differential 1-form $%
dG^{\prime }$ when expressed in terms of $\left( q^{\prime 1},q^{\prime
2},p_{1}^{^{\prime }},p_{2}^{\prime },p_{\phi ^{\prime }}\right) $ takes the
canonical form
\begin{equation}
dG^{\prime }=p_{1}^{^{\prime }}dq^{\prime 1}+p_{2}^{\prime }dq^{\prime 2},
\label{canonical reduction}
\end{equation}%
(\textit{canonical gyrokinetic reduction}) the variables $\mathbf{z}^{\prime
}\mathbf{=}\left( q^{\prime 1},q^{\prime 2},p_{1}^{^{\prime }},p_{2}^{\prime
},p_{\phi ^{\prime }},\phi ^{\prime }\right) $ are manifestly \textit{%
canonical gyrokinetic variables. }In fact, it is immediate to prove that the
Euler-Lagrange equations corresponding to the variational differential form (%
\ref{Lagrangan 1-form}) expressed in the variables $\mathbf{z}^{\prime }$
are canonical.

Particle dynamics expressed in terms of the canonical variables $\mathbf{z}%
^{\prime }$ denotes the so-called \textit{canonical gyrokinetic treatment}
(CGKT). The explicit construction of these variables has been first pointed
out by Littlejohn \cite{Littlejohn1983}, adopting the so-called Darboux
reduction method, by considering the vector potential $\mathbf{A,}$ and
hence the associated magnetic field $\mathbf{B}$ (equilibrium magnetic
field), as stationary. However, the proof \ - achieved in this way - of the
local existence of the diffeomorphism (\ref{canonical gyrokinetic variables}%
) and hence of the gyrokinetic canonical variables defined above $\mathbf{z}%
^{\prime },$\ is not generally applicable to general situations. In fact, to
reach it in Ref. \cite{Littlejohn1983} it was assumed\textit{\ that the
magnetic field admits, at least locally }(in configuration space)\textit{, a
family of nested toroidal magnetic surfaces}.

This raises, therefore, the issue of the general validity of such a
conclusion. In fact, the question is whether it applies only in the case of
equilibrium magnetic fields which are symmetric, i.e., which possess at
least one ignorable coordinate, or - at most - exhibit suitably small
deviations from a symmetric equilibrium. In fact, it is well known that the
proof of existence of smooth MHD equilibria with good magnetic surfaces
(namely which admit a family of locally nested toroidal magnetic surfaces in
a finite subset of configuration space) can only be achieved for symmetric
equilibria (1) or at most for magnetic fields which are asymptotically
close, in some sense, to equilibria of this type (2). As an example, in Ref.
\cite{White 1990} to obtain the canonical variables with the Darboux
reduction method, consistent with the requirement (2), it was assumed a
magnetic fields almost axi-symmetric, i.e., allowing actually only
infinitesimally small deviations from axi-symmetric toroidal geometry.

\section{On the existence of canonical gyrokinetic variables}

For definiteness let us pose, in this Section, the problem of the existence
of the canonical gyrokinetic variables in the framework of the exact
gyrokinetic formulation (Approach A). In order to solve the related problem
let us analyze the conditions of validity of the canonical gyrokinetic
reduction (\ref{canonical reduction}) in the particular case in which there
results $\mathbf{a=a(y}^{\prime })$ in $dG^{\prime }$ [see Eq.(\ref{exchange
term})]. For this purpose let us seek a diffeomorphism
\begin{equation}
{\mathbf{r}}^{\prime }\rightarrow \mathbf{q}^{\prime }(\mathbf{y}^{\prime }),%
\mathbf{q}^{\prime }\equiv (q^{\prime 1},q^{\prime 2},q^{\prime 3})
\label{diff-1}
\end{equation}%
denoting in principle arbitrary real and gyrokinetic variables. These can be
defined, in particular, in such a way that
\begin{equation}
\left( \mathbf{r}^{\prime },u^{\prime },p_{\phi ^{\prime }},\phi ^{\prime
}\right) \rightarrow \left( q^{\prime 1},q^{\prime 2},q^{\prime 3},\xi
^{\prime },p_{\phi ^{\prime }},\phi ^{\prime }\right) ,  \label{diff-2}
\end{equation}%
with $\xi ^{\prime }$ to be suitably defined, is a phase-space
diffeomorphism. Hence it follows that the differential 1-form $dG^{\prime }$
has necessarily the general form%
\begin{equation}
dG^{\prime }=f_{i}^{\prime }dq^{\prime i}\equiv dG"  \label{4}
\end{equation}%
where $f_{i}^{\prime }=f_{i}^{\prime }\left( \mathbf{q}^{\prime },\xi
^{\prime },p_{\phi ^{\prime }}\right) .$ The analysis of the conditions of
validity of the dynamical reduction - under which the differential 1-form $%
dG",$ as given by Eq.(\ref{4}), can be brought to its canonical form (\ref%
{canonical reduction}) - is straightforward. Let us first establish the
following lemma

\textbf{Lemma - Reduced form for}\emph{\ }$dG"$

\emph{Let us assume that the real functions }$f_{i}^{\prime }=f_{i}^{\prime
}\left( \mathbf{q}^{\prime },\xi ^{\prime },p_{\phi ^{\prime }}\right) $
\emph{(for }$i=1,2,3$)\emph{:}

\emph{1) are suitably smooth (i.e., at least C}$^{(2)}$\emph{);}

\emph{2) the set of gyrokinetic variables }$\left( q^{\prime 1},q^{\prime
2},f_{1}{}^{\prime },f_{2}{}^{\prime },p_{\phi ^{\prime }},\phi ^{\prime
}\right) $ \emph{are defined so that they all independent;}

\emph{3) are defined so that for at least an index }$i$\emph{\ (for }$%
i=1,2,3 $\emph{)} \emph{there results}
\begin{equation}
\frac{\partial f_{i}^{\prime }\left( \mathbf{q}^{\prime },\xi ^{\prime
},p_{\phi ^{\prime }}\right) }{\partial \xi ^{\prime }}=0
\end{equation}%
\emph{only in isolated points of the gyrokinetic phase-space} \emph{spanned
by the vector} $(\mathbf{q}^{\prime },\xi ^{\prime },p_{\phi ^{\prime }}).$%
\emph{\ }

\emph{Then a necessary and sufficient condition that the differential 1-form
}$dG"\equiv f_{i}^{\prime }dq^{\prime i}$\emph{\ can be represented in the
reduced form}%
\begin{equation}
dG"=f_{1}^{\prime }dq^{\prime 1}+f_{2}^{\prime }dq^{\prime 2}
\label{lemma-1}
\end{equation}
\emph{is that }$f_{3}^{\prime }=f_{3}^{\prime }\left( \mathbf{q}^{\prime
},\xi ^{\prime },p_{\phi ^{\prime }}\right) $ \emph{is a first integral of
motion, i.e., there results}%
\begin{equation}
df_{3}^{\prime }\left( \mathbf{q}^{\prime },\xi ^{\prime },p_{\phi ^{\prime
}}\right) =0.  \label{constant of motion}
\end{equation}

PROOF

Both the necessary and sufficient conditions are trivial. In fact, if $%
f_{3}^{\prime }$ is a first integral, since the Lagrangian 1-form is defined
up to an arbitrary gauge it follows%
\begin{equation}
f_{2}^{\prime }dq^{\prime 2}=d(f_{2}^{\prime }q^{\prime 2})-q_{3}^{\prime
}df_{3}^{\prime }\left( \mathbf{q}^{\prime },\xi ^{\prime },p_{\phi ^{\prime
}}\right) =0.
\end{equation}
On the other hand, if up to an arbitrary gauge transformation, the equation $%
f_{2}^{\prime }dq^{\prime 2}=0$ holds identically in a finite subset
[neighborhood] of gyrokinetic phase-space$,$ it follows necessarily Eq.(\ref%
{constant of motion}).

Provided the hypotheses of the lemma hold the following theorem has the
flavor of:

\textbf{Theorem 1 - Existence of canonical gyrokinetic variables}

\emph{In validity of the hypotheses of the Lemma, provided the gyrokinetic
transformation}
\begin{equation}
\left( \mathbf{r}^{\prime },u^{\prime },p_{\phi ^{\prime }},\phi ^{\prime
}\right) \rightarrow \left( q^{\prime 1},q^{\prime 2},f_{1}{}^{\prime
},f_{2}{}^{\prime },p_{\phi ^{\prime }},\phi ^{\prime }\right)
\end{equation}%
\emph{is a }$C^{(2)}$\emph{-diffeomorphism, it follows that:}

\emph{A) it is always possible to identify }%
\begin{eqnarray}
p_{1}^{\prime } &=&f_{1}{}^{\prime }, \\
p_{2}^{\prime } &=&f_{2}{}^{\prime }.
\end{eqnarray}%
\emph{in Eq.(\ref{lemma-1});}

\emph{B) the transformed variables }$\left( q^{\prime 1},q^{\prime
2},p_{1}^{\prime }=f_{1}{}^{\prime },p_{2}^{\prime }=f_{2}{}^{\prime
},p_{\phi ^{\prime }},\phi ^{\prime }\right) $ \emph{are canonical
gyrokinetic variables. }

PROOF

To prove the theorem one has to realize, first, that the
assumptions of the Lemma are indeed satisfied by the gyrokinetic
Lagrangian defined by Eq.\ref{Lagrangan 1-form}. Then the proof is
an immediate consequence of the Lemma.

A basic consequence of the theorem here pointed out is that the
adoption of canonical gyrokinetic variables in gyrokinetic theory
is only permitted if the gyrokinetic Lagrangian, besides $\phi
^{\prime }$, has an additional ignorable coordinate, $q^{\prime
3}$ and hence it admits necessarily two first integrals of motion
$p_{\phi }^{\prime }$ and $p_{3}^{\prime }$. In turn, one can show
that this condition implies that both the electric and
magnetic fields (as well the corresponding EM potentials $\Phi ,\ \mathbf{A}$%
) must be symmetric \cite{Tessarotto1996}. \emph{This implies that
if the equilibrium magnetic field\ }$B$\emph{\ is non-symmetric,
or more generally is locally chaotic (i.e., it\ does not admit
locally a family of nested magnetic surfaces), the gyrokinetic
transformation (\ref{canonical gyrokinetic variables}) - in the
sense of approach A - does not exist.}

\section{Discussion and conclusions}

In this paper the conditions of existence of the canonical gyrokinetic
variables for a classical charged point-particle have been investigated. We
have shown that - in the framework of an exact gyrokinetic treatment
(Approach A) - these variables can only be achieved provided the particle
gyrokinetic Lagrangian is symmetric. This means, actually, that it must have
generally \textit{two} ignorable coordinates ($\phi ^{\prime }$ and $%
q^{\prime 3}$).

The extension of these results to the asymptotic gyrokinetic treatment
(Approach B) is non-trivial. In fact, even small perturbations of the EM
field can in principle produce significant local (and even non-local)
stochastic effects. Nevertheless, near an axi-symmetric MHD equilibrium,
i.e., for magnetic fields which are weakly non-symmetric (and
weakly-turbulent) - in the sense that they are characterized by suitably
small deviations from a symmetric equilibrium - one should expect CGKT to
hold locally, at least, in an asymptotic sense, a result earlier pointed out
by White \cite{White 1990}.

However, these conclusions cannot be extended to general situations. As an
example, in Stellarators magnetic surfaces may only exist locally namely in
the neighborhood of nested magnetic surfaces only. Therefore it would be
highly desirable to be able to construct gyrokinetic canonical variables
which result independent of the magnetic field geometry and apply also to
the case of chaotic magnetic fields. An example is given by so-called
quasi-symmetric \cite{Tessarotto 1995} MHD equilibria which arise in
Stellarators. These equilibria - which actually may be strongly
non-symmetric - are expected to be characterized, at most, by a family of
isolated nested toroidal magnetic surface. Typically, in the intermediate
regions between these surfaces the magnetic field is chaotic. Another
typical situation is that arising in the presence of local \ MHD/kinetic
turbulence, in which EM perturbations may give rise to local chaotic
behavior of the magnetic field. These results are potentially relevant for
their implications for theoretical investigations and numerical simulations
of magnetized plasmas. In fact, the regularity conditions on the EM fields,
to be imposed for the validity of CGKT, may be locally violated in typical
MHD equilibria. For example, a consistent kinetic description or a numerical
gyrokinetic particle simulation of a magneto-plasma in these variables
cannot be achieved unless\textit{\ the EM field is weakly non-symmetric in
whole the domain }occupied by the plasma. \

It should be stressed that there is a simple alternative to the description
based on canonical gyrokinetic variables. This is represented by the
super-abundant canonical gyrokinetic treatment \emph{(super-abundant CGKT)}
pointed out in Ref.\cite{Tessarotto2004}, which preserves both the
Hamiltonian character of the equations and - unlike CGKT - is applicable
also in the presence of a chaotic magnetic field.\emph{\ \ }Basic features
of this approach are in fact that: 1) no symmetry (or quasi-symmetry)
assumption is required for the magnetic field, so that it holds also in the
case of chaotic magnetic fields ; 2) the Hamilton equations for the
canonical pair ( $\mathbf{r}^{\prime },p_{\mathbf{r}^{\prime }}$) are in
vector form. Its formulation is summarized by the following constrained
variational principle,

\textbf{Theorem 2 - Superabundant CGKT}

\emph{Let} $\mathbf{x=}\left( \mathbf{r,}p_{\mathbf{r}}\right) $\emph{\ be
the canonical state of a charged point particle described by the Hamiltonian
}%
\begin{equation}
H(\mathbf{r,}\ p\mathbf{_{r}},\ t)=\frac{1}{2m}\left[ p_{\mathbf{r}}-\frac{q%
}{\varepsilon c}\mathbf{A}\right] ^{2}+\frac{q}{\varepsilon }\Phi
\label{HAM}
\end{equation}%
\emph{and introduce the diffeomorphism}%
\begin{equation}
\mathbf{x=}\left( \mathbf{r,}p_{\mathbf{r}}\right) \rightarrow \mathbf{x}%
^{\prime }\mathbf{=}\left( \mathbf{r}^{\prime }\mathbf{,}p_{\mathbf{r}%
^{\prime }},\phi ^{\prime },p_{\phi ^{\prime }}\right) ,
\end{equation}%
\emph{\ } \emph{(}$\mathbf{x}^{\prime }\equiv $\emph{superabundant canonical
gyrokinetic state), where}
\begin{equation}
p_{\mathbf{r}^{\prime }}\ \equiv \frac{q}{\varepsilon c}\mathbf{A}^{\ast }=%
\frac{\partial \mathcal{L}^{\prime }}{\partial \left( \frac{d}{dt}\mathbf{r}%
^{\prime }\right) }\equiv m\mathbf{v}^{\prime }\mathbf{+}\frac{q}{%
\varepsilon c}\mathbf{A}^{\ast },  \label{momento canonico girocinetico}
\end{equation}%
\emph{and the gyrokinetic Hamiltonian (\ref{GKT HAM}) is represented in the
form}
\begin{equation}
K(\mathbf{x}^{\prime },t)=-p_{\phi ^{\prime }}\Omega ^{\prime }+\frac{1}{2m}%
\left[ p_{\mathbf{r}^{\prime }}\ -\frac{q}{\varepsilon c}\mathbf{A}^{\ast }(%
\mathbf{r}^{\prime }\mathbf{,}u^{\prime },p_{\phi ^{\prime }},t)\right] ^{2}+%
\frac{q}{\varepsilon }\Phi ^{\ast }\mathbf{r}^{\prime }\mathbf{,}u^{\prime
},p_{\phi ^{\prime }},t).
\end{equation}%
\emph{It follows that: 1) }$\mathbf{x}^{\prime }(t)$ \emph{is the extremal
curve of the functional }$S(\mathbf{x}^{\prime
})=\int\limits_{t_{1}}^{t_{2}}dt\left\{ \overset{\cdot }{\mathbf{r}}%
^{\prime }\cdot p_{\mathbf{r}^{\prime }}-\overset{\cdot }{{\phi }}^{\prime
}p_{\phi }^{\prime }-K\right\} $ \emph{which satisfies the synchronous
variational principle }$\delta S(\mathbf{x}^{\prime })=0,$ \emph{with the
variations} $\delta \mathbf{x}^{\prime }\equiv \left( \delta \mathbf{r}%
^{\prime }\mathbf{,}\delta p_{\mathbf{r}^{\prime }},\delta \phi ^{\prime
},\delta p_{\phi ^{\prime }}\right) $ \emph{to be taken as linearly
independent and the function }$u^{\prime }(t)$\emph{\ to be considered
extremal with respect to }$\delta p_{\phi ^{\prime }}$ \emph{and} $\delta
\mathbf{r}^{\prime }$\emph{, i.e., such that there results identically }%
\begin{eqnarray}
\delta p_{\mathbf{r}^{\prime }}\cdot \frac{\partial u^{\prime }}{\partial p_{%
\mathbf{r}^{\prime }}} &\equiv &0; \\
\delta \mathbf{r}^{\prime }\cdot \frac{\partial u^{\prime }}{\partial
\mathbf{r}^{\prime }} &=&0;
\end{eqnarray}%
\emph{2) }$\mathbf{x}^{\prime }(t)$ \emph{is canonical with respect to the
gyrokinetic Hamiltonian\ }

\emph{PROOF}

The proof is straightforward. In particular, by taking the variations with
respect to $p_{\mathbf{r}^{\prime }}$ and $\mathbf{r}^{\prime },$ the
Euler-Lagrange equations for $\mathbf{r}^{\prime }$ and $p_{\mathbf{r}%
^{\prime }}$ are simply
\begin{equation}
\frac{d}{dt}\mathbf{r}^{\prime }=\frac{\partial }{\partial p_{\mathbf{r}%
^{\prime }}}K(\mathbf{x}^{\prime },t)=\frac{1}{m}\left[ p_{\mathbf{r}%
^{\prime }}\ -\frac{q}{\varepsilon c}\mathbf{A}^{\ast }\right] .
\label{rHeqns}
\end{equation}%
\begin{equation}
\frac{d}{dt}p_{\mathbf{r}^{\prime }}=-\frac{\partial }{\partial \mathbf{r}%
^{\prime }}K(\mathbf{x}^{\prime },t).  \label{prHeqns}
\end{equation}%
Finally the equations for $p_{\phi }^{\prime }$ and $\phi ^{\prime }$ are
manifestly
\begin{equation}
\frac{d}{dt}p_{\phi }^{\prime }=-\frac{\partial }{\partial \phi ^{\prime }}K(%
\mathbf{x}^{\prime },t)=0,  \label{pphiHequns}
\end{equation}%
\begin{equation}
\frac{d}{dt}\phi ^{\prime }=\frac{\partial }{\partial p_{\phi }^{\prime }}K(%
\mathbf{x}^{\prime },t).  \label{phiHeqns}
\end{equation}%
It is immediate to prove that these equations coincide with the equations of
motion obtained from the Lagrangian (\ref{GKT HAM}).


\section*{Acknowledgments}
Work developed in cooperation with the CMFD Team, Consortium for
Magneto-fluid-dynamics (Trieste University, Trieste, Italy). \
Research developed in the framework of the MIUR (Italian Ministry
of University and Research) PRIN Programme: \textit{Modelli della
teoria cinetica matematica nello studio dei sistemi complessi
nelle scienze applicate}. The support COST Action P17 (EPM,
\textit{Electromagnetic Processing of Materials}) and GNFM
(National Group of Mathematical Physics) of INDAM (Italian
National Institute for Advanced Mathematics) is acknowledged.

\section*{Notice}
$^{\S }$ contributed paper at RGD26 (Kyoto, Japan, July 2008).

\end{document}